**Astro2020 Science White Paper**

**'Auxiliary' Science with the WFIRST Microlensing Survey: Measurement of the Compact Object Mass Function over Ten Orders of Magnitude; Detection of ~$10^5$ Transiting Planets; Astroseismology of ~$10^6$ Bulge Giants; Detection of ~$5\times10^3$ Trans-Neptunian Objects; and Parallaxes and Proper Motions of ~$6\times10^6$ Bulge and Disk Stars.**

**Thematic Areas:**   ☒ Planetary Systems   ☐ Star and Planet Formation
☒ Formation and Evolution of Compact Objects   ☐ Cosmology and Fundamental Physics
☒ Stars and Stellar Evolution   ☒ Resolved Stellar Populations and their Environments
☒ Galaxy Evolution   ☐ Multi-Messenger Astronomy and Astrophysics

**Principal Author:**
Name: B. Scott Gaudi
Institution: The Ohio State University (OSU)
Email: gaudi.1@osu.edu
Phone: 614-292-1914

**Co-authors:**

Rachel Akeson (Caltech/IPAC; rla@ipac.caltech.edu), Jay Anderson (STScI; jayander@stsci.edu), Etienne Bachelet (Las Cumbres Observatory; etibachelet@gmail.com), David P. Bennett (NASA Goddard Space Flight Center and University of Maryland; david.p.bennett@nasa.gov), Aparna Bhattacharya (NASA Goddard Space Flight Center; abhatta5@umd.edu), Valerio Bozza (Università di Salerno; valboz@sa.infn.it), Sebastiano Calchi Novati (Caltech/IPAC; snovati@ipac.caltech.edu), Calen B. Henderson (Caltech/IPAC; chenderson@ipac.caltech.edu), Samson A. Johnson (OSU; johnson.7080@osu.edu), Jeffrey Kruk (NASA Goddard Space Flight Center; jeffrey.w.kruk@nasa.gov), Jessica R. Lu (UC Berkeley; jlu.astro@berkeley.edu), Shude Mao (Tsinghua University; shude.mao@gmail.com), Benjamin T. Montet (University of Chicago; bmontet@uchicago.edu), David M. Nataf (The Johns Hopkins University; dnataf1@jhu.edu), Matthew T. Penny (OSU; penny.24@osu.edu), Radoslaw Poleski (The Ohio State University; radek.poleski@gmail.com), Clément Ranc (NASA Goddard Space Flight Center; clement.ranc@nasa.gov), Kailash Sahu (STScI; ksahu@stsci.edu), Yossi Shvartzvald (Caltech/IPAC; yossishv@gmail.com), David N. Spergel (Princeton; dns@astro.princeton.edu), Daisuke Suzuki (ISAS/JAXA; dsuzuki@ir.isas.jaxa.jp), Keivan G. Stassun (Vanderbilt University; keivan.stassun@vanderbilt.edu), Rachel A. Street (Las Cumbres Observatory; rstreet@lco.global)

**Abstract**:
The Wide Field Infrared Survey Telescope (WFIRST) will monitor ~2 deg$^2$ toward the Galactic bulge in a wide (~1-2 μm) W149 filter at 15-minute cadence with exposure times of ~50s for 6 seasons of 72 days each, for a total ~41,000 exposures taken over ~432 days, spread over the 5-year prime mission. This will be one of the deepest exposures of the sky ever taken, reaching a photon-noise photometric precision of 0.01 mag per exposure and collecting a total of ~$10^9$ photons over the course of the survey for a $W149_{AB} \sim 21$ star. Of order $4 \times 10^7$ stars will be monitored with $W149_{AB} < 21$, and $10^8$ stars with $W145_{AB} < 23$. The WFIRST microlensing survey will detect ~54,000 microlensing events, of which roughly 1% (~500) will be due to isolated black holes, and ~3% (1600) will be due to isolated neutron stars. It will be sensitive to (effectively) isolated compact objects with masses as low as the mass of Pluto, thereby enabling a measurement of the compact object mass function over 10 orders of magnitude. Assuming photon-noise limited precision, it will detect ~$10^5$ transiting planets with sizes as small as ~2 $R_{Earth}$, perform asteroseismology of ~$10^6$ giant stars, measure the proper motions to ~0.3% and parallaxes to ~10% for the $6 \times 10^6$ disk and bulge stars in the survey area, and directly detect ~$5 \times 10^3$ Trans-Neptunian objects (TNOs) with diameters down to ~10 km, as well as detect ~$10^3$ occulations of stars by TNOs during the survey. All of this science will completely serendipitous, i.e., it will not require modifications of the WFIRST optimal microlensing survey design. Allowing for some minor deviation from the optimal design, such as monitoring the very center of our Galaxy, would enable an even broader range of transformational science.

We endorse the findings and recommendations published in the National Academy reports on Exoplanet Science Strategy and Astrobiology Strategy for the Search for Life in the Universe. This white paper extends and complements the material presented therein. In particular, this white paper supports the recommendation of the National Academy Exoplanet Science Strategy report that "NASA should launch WFIRST to conduct its microlensing survey of distant planets and to demonstrate the technique of coronagraphic spectroscopy on exoplanet targets." (National Academies of Sciences, Engineering, and Medicine, *Exoplanet Science Strategy*)

**The Wide Field Infrared Survey Telescope: 100 Hubbles for the 2020s.**

The Wide Field Infrared Survey Telescope (WFIRST) is a 2.4m space telescope with a 0.281 deg$^2$ field of view for near-IR imaging and slitless spectroscopy and a coronagraph designed for $> 10^8$ starlight suppression (Akeson et al. 2019). While WFIRST does not have the UV imaging/spectroscopic capabilities of the Hubble Space Telescope, for wide field near-IR surveys WFIRST is hundreds of times more efficient. In this white paper, we demonstrate how powerful this capability is, by noting the transformational science that will be possible as a byproduct of one of the primary surveys planned for the mission: the microlensing survey for exoplanets in the Galactic bulge.

**Auxiliary Science with the WFIRST Microlensing Survey**

The WFIRST microlensing survey toward the Galactic bulge (Penny et al. 2019a) will potentially enable a very broad range of astrophysics, well beyond the primary goal of completing the census of cold exoplanetary systems. This enormous additional potential of the survey follows from the large number of stars being surveyed, the large number of measurements that will be made for each these stars, and the quality of the photometric and astrometric precision that will be achieved for these stars, which will range from good to exquisite, depending on the brightness of the star and the level to which systematics can be controlled.

In this white paper, we provide a sampling of the range of the science that will be possible, including auxiliary science from the microlensing events themselves, additional exoplanet science, stellar and remnant astrophysics, Galactic structure, and solar system science.

There are a few things to note. First, this overview contains an admixture of potential applications, some of which have been worked out in some detail, and some of which have merely been suggested, with no real effort made to demonstrate their viability. These latter topics provide an important line of future work to vet their applicability. Second, we are primarily concentrating on auxiliary science that can potentially be extracted without changing the microlensing survey specifications. However, some of these applications will require new data reduction algorithms, or precursor or follow-up observations, in order to realize their full science potential. Finally, we note that several of these applications require excellent-to-exquisite control of systematics in the photometric and astrometric measurements. The extent to which systematics can be controlled at the levels required is unclear. However, we note that it is likely that the microlensing survey itself will provide the best opportunity to characterize and remove systematics, by its very nature of requiring a large number of observations of a very large number of point sources. We discuss this in a bit more detail below.

**Properties of the WFIRST Microlensing Survey**

We begin by summarizing the properties of the WFIRST microlensing survey (for more details, see Penny et al. 2019a). As currently envisioned, the survey will consist of seven contiguous fields between Galactic longitudes of roughly -0.5 and 1.5 deg, and Galactic latitudes of -0.5 and -2 deg, for a total area of 1.97 deg$^2$. The fields will be imaged in six 72-day campaigns spread over the nominal five-year mission lifetime, resulting in a total survey time of 432 days. During each campaign, each field will be observed every 15 minutes in a wide (W149) filter that spans 0.93-2 microns, and once every 12 hours in a bluer filter (likely Z087 that spans 0.76-0.98 microns). The exposure time per epoch in these filters will be 46.8s for W149 and 286s for Z087. Because of the layout of focal plane, only ~90% of the area will be covered for the full time, while the remaining ~10% will be observed only in either the spring or fall seasons. For stars in the primary area, there will be a total of ~41,000 epochs in W149 and ~900 epochs in Z087. The single measurement photon noise limited photometric and astrometric precisions will be ~0.8% and ~0.5 mas for the 6 million stars with $H_{AB}$<19, and ~1% and 1.5 mas for the 40 million stars with $H_{AB}$ < 21. WFIRST will collect roughly one billion photons for a star with $H_{AB}$=21 in the wide filter over the course of the mission.

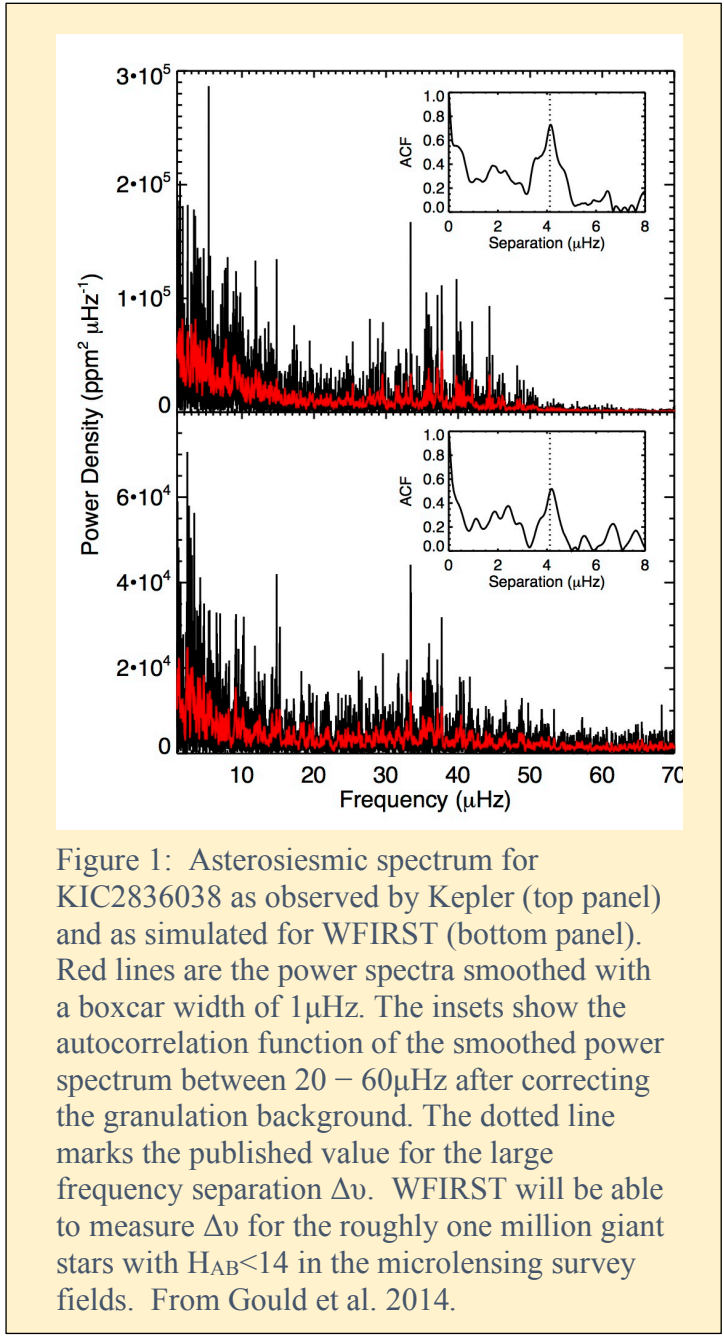

Figure 1: Asterosiesmic spectrum for KIC2836038 as observed by Kepler (top panel) and as simulated for WFIRST (bottom panel). Red lines are the power spectra smoothed with a boxcar width of 1µHz. The insets show the autocorrelation function of the smoothed power spectrum between 20 − 60µHz after correcting the granulation background. The dotted line marks the published value for the large frequency separation $\Delta\upsilon$. WFIRST will be able to measure $\Delta\upsilon$ for the roughly one million giant stars with $H_{AB}$<14 in the microlensing survey fields. From Gould et al. 2014.

**Microlensing Auxiliary Science**

The microlensing survey will be sensitive to microlensing due to isolated compact objects with masses from that of Pluto to roughly 30 times the mass of the Sun. The upper limit is set by the typical timescale of a microlensing event due to a ~30 solar mass compact object lens relative to the lifetime of the primary mission. The lower limit is due to the small amplitude and low signal-to-noise ratio of microlensing by very low-mass objects. We expect 54,000 microlensing events (with impact parameter <3 Einstein ring radii) due to known populations alone, e.g., stars and stellar remnants (Penny et al. 2019a). Thus, it will be able to measure or constrain the mass function of compact objects over roughly 10 orders of magnitude in mass. This includes free-floating planets, brown dwarfs, stars, and stellar remnants (see also Lu et al. 2019). For the upper end of this range, it will be possible to measure the masses of the objects directly from higher-order microlensing effects (astrometric microlensing) from the survey data itself (Gould & Yee 2014, Lu et al. 2016), whereas for lower-mass objects, either additional measurements with ground-based extremely large telescopes equipped with adaptive optics may be required, or the mass function will only be determined statistically (Lu et al. 2019). It will also be possible to probe the binarity of these compact objects, including the companion frequency and mass ratio and separation distribution, for binaries with separations within roughly a decade of the Einstein ring radius of the primary. Finally, it will also be possible to detect tertiaries in some cases.

With the large number of microlensing events, it will be possible to measure the microlenisng optical depth and event rate over the ~2 square degrees of the target fields. The optical depth, in particular, is proportional to the integral of the total mass density of compact objects along the line of sight, and as such is a very sensitive probe of the mass distribution of the Galactic disk and the Galactic bulge (Griest et al. 1991). The optical depth can be used to probe the total mass, shape, and orientation of the Galactic bar (Kiraga & Paczynski 1994; Zhao et al. 1995), and the dark matter profile near the Galactic center (Wegg et al. 2016). The event rate is additionally sensitive to the velocity distribution and mass function of the compact objects in the stellar populations, and will enable a measurement of the bulge initial mass function (Wegg et al. 2017)

Roughly 5% of all microlensing events exhibit clear signatures of caustic crossings due to the lens being composed of a nearly equal mass binary with a projected separation near the Einstein ring radius. With the near-continuous sampling and 15-minute cadence, nearly all of the caustic crossings of giant sources (which last ~10 hours) will be well-resolved, thus allowing one to measure the limb-darkening for these stars in the wide filter (Albrow et al. 1999).

**Stellar Astrophysics and Resolved Stellar Populations**

With a typical total exposure time of $~2 \times 10^6$ s per field in the wide (W149) filter and $~3 \times 10^5$ s per field in the blue (Z087) filter, it will be possible to construct a very deep color-magnitude diagram over the ~2 square degrees targeted by the microlensing survey. The limiting magnitude will likely be set by confusion, rather than photon noise. Nevertheless, it should be possible to measure the luminosity function of stars in the Galactic bulge down to nearly the bottom of the main sequence, as well as identify unusual stellar populations down to quite faint magnitudes, including potentially young stars, blue stragglers, stellar clusters, cataclysmic variables, and white dwarfs. Such studies will be enhanced by the fact that it will be possible to measure the proper motion and potentially parallax for a significant subset of the sources in the bulge field, thus allowing for discrimination between bulge and disk populations. This science would be significantly enhanced by observations in one or more bluer filters at similar angular resolution, either with WFIRST itself, or with Hubble Space Telescope or Euclid precursor observations (Penny et al. 2019b).

Of course, it will be possible to identify many variable sources in the fields, over a broad range of amplitudes and time scales. Variables can be identified with amplitudes greater than a few millimagnitudes, with time scales from tens of minutes up to the ~5-year duration of the mission. This will allow for the identification of a large number of stellar flares, eruptive variables, and pulsating variables, and even rear, currently unknown types of stellar variability, thereby enabling a wide variety of stellar astrophysics.

From Kepler, it is known that ~3% of sunlike vary at the 1% level (Montet et al. 2017). The bulge would provide a laboratory to determine how metallicity affects the nature and frequency of variability of this amplitude in sunlike stars (Karoff et al. 2018; Witzke et al. 2018).

Perhaps most exciting, however, is the potential for WFIRST to do asteroseismology on, and thus determine the mass and radii of, a significant number of stars in the survey area. Gould et al. (2015) estimate that WFIRST will obtain asteroseismic information on the roughly $10^6$ giant stars with $W149_{AB} < 17$ in the target fields (see Figure 1). They also estimate that WFIRST will obtain precise (~0.3%) distances to these stars via parallax. This is the only way to measure masses and radii of a large number of bulge giants, which are postulated to have abundances that are substantially different than local populations. Parallaxes are particularly important, as astrosiesmic masses and radii generally come from scaling relations calibrated to the Sun, which may not apply to the α- and possibly He- enhanced bulge stars (Nataf et al. 2011).

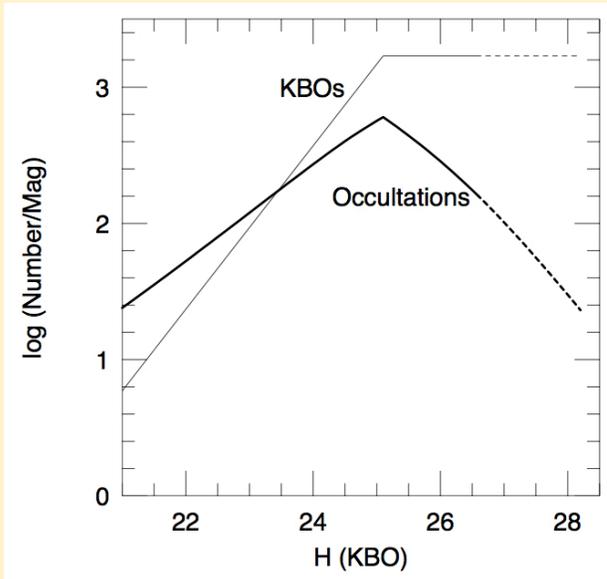

Figure 2. The expected number of Trans-Neptunian Object (or KBO) detections (light solid line) and TNO occultations (heavy solid line) per magnitude that can be detected from the WFIRST microlensing survey. Magnitudes here are in the Vega system, which is related to the AB system used in the text by $H_{Vega} = H_{AB} - 1.39$. WFIRST will detect TNOs significantly fainter the current survey limits of $H_{Vega} = 26$. KBOs with $H_{Vega} < 23.5$ will occult more than one star during the survey, potentially allowing for crude constraints on their shape. From Gould 2014.

**Galactic Structure**

WFIRST can potentially obtain photon-noise limited parallaxes of <10% and proper motion measurements of <0.3% (0.01 mas/yr) for the ~38 million bulge and disk stars with $W149_{AB}<21$ in its field of view. When combined with multicolor photometry in at least one bluer filter (to complement the measurements in W149 and Z087), it will be possible to estimate the effective temperature, metallicity, age, luminosity, and foreground extinction for all of these stars. This is similar to the number of measurements than will be achievable with *Gaia*, although of course the population of stars will be very different and complementary to that obtained by *Gaia*. From this dataset, it will be possible to extract an enormous amount of Galactic structure science, including a determination of the bulge mass and velocity distribution (including bar structure), the stellar density and velocity distribution of the Galactic disk, the metallicity and age distribution of the disk and bulge, and the three-dimensional distribution of dust along the line of sight toward the bulge fields.

**Solar System Science**

Gould (2014) estimates that WFIRST will detect ~5000 Trans-Neptunian objects (TNOs) down to absolute magnitudes of $W149_{AB}$~30 (corresponding to diameters of D~10 km) over ~17 deg$^2$, enabling a precise determination of the size distribution of TNOs down to, and substantially below, the collisional break at ~100 km (see Figure 2). Furthermore, the orbital elements of these objects will be measured to fractional precisions of a few percent, allowing for dynamical classification into the canonical classical, resonant, and scattered populations, as well as identification of objects with unique or unusual orbits. Binary companions to these objects will be discovered down to fainter magnitudes of $H_{AB}$~30.4, corresponding to diameters of 7 km, thus allowing for the determination of the binary size and separation distribution as a function of dynamical class. Finally, there will be ~1000 occulations of stars due to these TNOs during the bulge survey (Figure 2), allowing for a statistical estimate of the size and albedo distributions of TNOs as a function of magnitude and dynamical class. The several dozen TNOs discovered with $H_{AB}$<25 (D>100 km) will occult multiple stars during the survey, allowing for crude measurement of the shape of the larger TNOs.

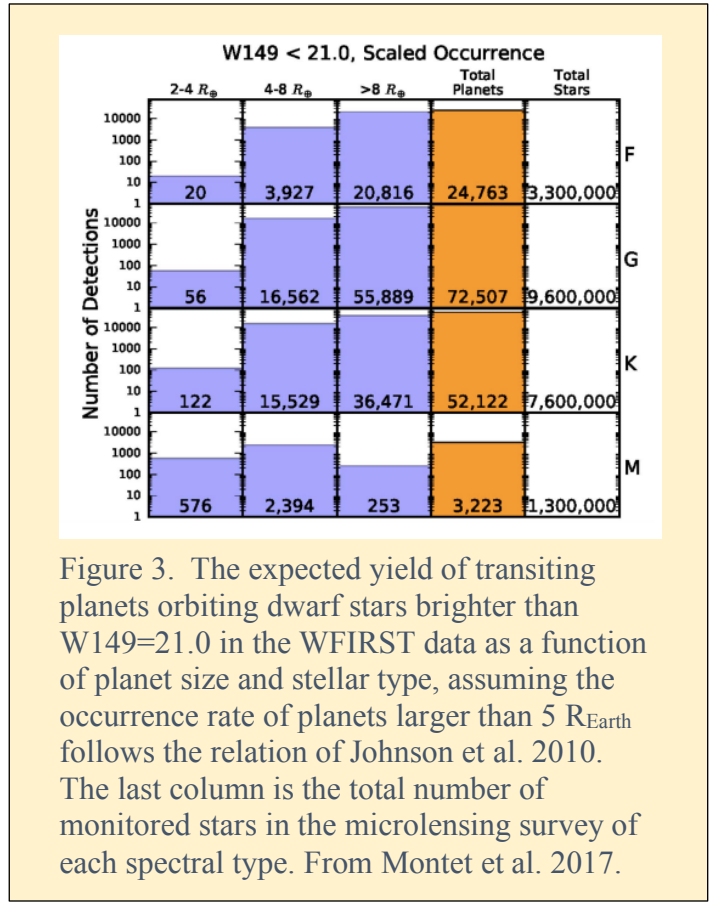

Figure 3. The expected yield of transiting planets orbiting dwarf stars brighter than W149=21.0 in the WFIRST data as a function of planet size and stellar type, assuming the occurrence rate of planets larger than 5 $R_{Earth}$ follows the relation of Johnson et al. 2010. The last column is the total number of monitored stars in the microlensing survey of each spectral type. From Montet et al. 2017.

### Transiting Planets in the Galactic Disk and Bulge

The WFIRST microlensing mission will measure precise light curves and relative parallaxes for millions of stars, giving it the potential to detect and characterize short-period transiting planets all along the line of sight through the Galactic disk and into the Galactic bulge (Bennett & Rhie 2002; Monet et al. 2017). These light curves will enable the detection of more than $10^5$ transiting planets with radii down to 2 $R_{Earth}$, whose host stars will have measured distances in most cases. Although most of these planets cannot be followed up, several thousand hot Jupiters can be confirmed directly by detection of their secondary eclipses in the WFIRST data. Additionally, some systems of small planets may be confirmed by detecting transit timing variations over the duration of the WFIRST microlensing survey (Montet et al. 2017). Finally, many more planets may be validated by ruling out potential false positives. The combination of WFIRST transits and microlensing will provide a complete picture of planetary system architectures, from the very shortest periods to unbound planets, as a function of Galactocentric distance.

### Detector Characterization: Recommendations

As mentioned previously, many of these applications will require new data reduction algorithms in order to realize their full science potential, as well as excellent-to-exquisite control of systematics in the photometric and astrometric measurements. The extent to which systematics can be controlled at the levels required is unclear. With its ~41,000 dithered images of ~$10^8$ point sources with known colors and nearly constant fluxes, the bulge survey will likely provide one of the best datasets with which to characterize WFIRST's wide-field imager. In particular, the bulge survey pipeline will be designed to measure the fluxes and positions of all of the stars in the field of view, while simultaneously measuring the response function of each of the wide field instrument (WFI) pixels, and how this response function varies with time. This time-variable response function will be used to calibrate detector artifacts such as persistence, inter-pixel capacitance, reciprocity failure, non-linearity, and intra-pixel response variations.